\documentstyle[twocolumn,floats,epsf,prb,aps]{revtex}

\newcommand{\beq}{\begin{equation}}
\newcommand{\eeq}{\end{equation}}
\newcommand{\bea}{\begin{eqnarray}}
\newcommand{\eea}{\end{eqnarray}}
\newcommand{\bi}{\begin{itemize}}
\newcommand{\ei}{\end{itemize}}
\newcommand{\be}{\begin{enumerate}}
\newcommand{\ee}{\end{enumerate}}
\newcommand{\bd}{\begin{description}}
\newcommand{\ed}{\end{description}}
\newcommand{\bfig}{\begin{figure}}
\newcommand{\efig}{\end{figure}}
\newcommand{\bfigu}{\begin{figure*}}
\newcommand{\efigu}{\end{figure*}}

\newcommand{\phrb}{Phys. Rev. B } 
 
\newcommand{\phrl}{Phys. Rev. Lett. }

\newcommand{\jm}{J. Magn. Magn. Mater. } 
  
\newcommand{\jap}{J. Appl. Phys. }

\begin{document} 
\twocolumn[\hsize\textwidth\columnwidth\hsize
\csname @twocolumnfalse\endcsname
\draft   
 
\title{Finite-Size and surface effects in maghemite nanoparticles: Monte Carlo 
simulations} 
\author{\`Oscar Iglesias and Am\'{\i}lcar Labarta} 
\address{Departament de F\'{\i}sica Fonamental, 
Universitat de Barcelona,  
Diagonal 647, 08028 Barcelona, Spain} 
\date{\today}   
\maketitle   
 
\begin{abstract} 
Finite-size and surface effects in fine particle systems are investigated
by Monte Carlo simulation of a model of a $\gamma$-Fe$_2$O$_3$ (maghemite)
single particle. Periodic boundary conditions have been used to simulate
the bulk properties and the results compared with those for a spherical 
shaped particle with free boundaries to evidence the role
played by the surface on the anomalous magnetic properties displayed by 
these systems at low temperatures. Several outcomes of the model are in 
qualitative agreement with the experimental findings. A reduction of the 
magnetic ordering temperature, spontaneous magnetization, and coercive field 
is observed as the particle size is decreased. Moreover, the hysteresis
loops become elongated with high values of the differential susceptibility,
resembling those from frustrated or disordered systems. These facts are
consequence of the formation of a surface layer with higher degree of magnetic
disorder than the core, which, for small sizes, dominates the 
magnetization processes of the particle. However, in contradiction with
the assumptions of some authors, our model does not predict the freezing of the 
surface layer into a spin-glass-like state. The results indicate that magnetic 
disorder at the surface simply facilitates the thermal demagnetization
of the particle at zero field, while the magnetization is increased at
moderate fields, since surface disorder diminishes ferrimagnetic correlations
within the particle. The change in shape of the hysteresis loops with the
particle size demonstrates that the reversal mode is strongly influenced
by the reduced atomic coordination and disorder at the surface.
\end{abstract} 
\date{\today}
]

\section{Introduction} 

The magnetic properties of fine particles are strongly influenced by finite-size
and surface effects, their relevance increasing as the particle size decreases.
Finite-size effects are due to the nanometric size of the particles, while surface
effects are related to the symmetry breaking of the crystal structure at the boundary
of each particle. These effects are manifested in nanometric particles through a
wide variety of anomalous magnetic properties with respect to those of bulk
materials.
The magnetic characterization of these systems has put forward the
controversial issue of distinguishing between the contributions coming from 
finite-size and surface effects to their peculiar magnetic properties. 
For instance, alternately explanations to the reduction of the saturation 
magnetization $M_s$ - a common experimental observation in fine particle systems - 
has been given in the past. 
Early models postulated the existence of a so-called
dead magnetic layer induced by the demagnetization of the surface spins, which
causes a reduction in $M_s$ because of its paramagnetic response. In more recent 
works devoted to the study of different ferrimagnetic oxides 
- $\gamma$-Fe$_2$O$_3$, NiFe$_2$O$_4$, CoFe$_2$O$_4$, CuFe$_2$O$_4$, in the form of 
nanometric particles \cite{Coeyprl71,Morrishjap81,Kodamaprl96,Linjm95,Jiangjpcm99} - a random 
spin canting at the surface, caused by competing antiferromagnetic (AF) interactions,
was observed by M\"{o}ssbauer spectroscopy \cite{Morrishjap81}, 
polarized \cite{Linjm95} and inelastic \cite{Gazeaueul97} neutron scattering, 
and ferromagnetic (FM) resonance \cite{Gazeaujm98}.
The origin of this non-collinear arrangement of the spins was discussed by several
authors supporting the surface \cite{Morrishjap81,Ochijpsj81,Okadajm83,Hanedajap88} 
or the finite-size explanations \cite{Pankhurstprl91,Parkerprb93,Linderothjap94,Kodamaprl97,Moralesjpcm97},
but up to the moment no clear conclusions have been established. 

All these ferrimagnetic fine particles share a singular phenomenology at low 
temperatures. Among the static properties, experiments have shown that 
the hysteresis loops display high closure fields and do not saturate 
\cite{Kodamaprl97,Garciaprb99,Martinezprl98}even at fields of 
the order of 50 T, which indicates that 
the anisotropy fields cannot be the only responsible mechanism for the 
magnetization reversal. 
Low magnetization as compared to bulk, shifted loops after field cooling 
and irreversibilities between the field cooling and zero field cooling processes 
even at high fields are also observed \cite{Jiangjpcm99,Garciaprb99,Martinezprl98}. 
Moreover, the existence of aging phenomena \cite{Jonssonprl95,Jonssonprb98} 
in the time-dependence of the magnetization, indicates that there must be 
some kind of freezing leading to a complex hierarchy of energy levels. 
Whether these phenomena can be ascribed to intrinsic properties 
of the particle itself (spin-glass state of the surface which creates an 
exchange field on the core of the particle \cite{Kodamaprl97,Martinezprl98}), 
or they are due to a collective behaviour induced by interparticle 
interactions \cite{Batlleprb97,Dormannjm98,Morupprl94}, 
has been the object of controversy \cite{Jonssonprb00}. 

Up to the moment there has been no model giving a clear-cut explanation of all the
above mentioned phenomenology, but some works addressing part of the issues have been 
published in recent years. 
The first atomic-scale model of the magnetic behaviour of individual ferrimagnetic 
nanoparticles is due to Kodama and Berkowitz \cite{Kodamaprb99}. The authors presented
results of calculations of a micromagnetic model of maghemite particles which were
based on an energy minimization procedure instead of the Monte Carlo (MC) method. 
They used Heisenberg spins with enhanced anisotropy at the surface with respect to the
core and included vacancies and broken bonds at the surface, arguing that these are 
indeed necessary to obtain hysteresis loops with enhanced coercivity and high-field
irreversibility. 
Later, Kachkachi et al. \cite{Kachkachieuj00,Kachkachijm00,Kachkachijpctb}  
performed MC simulations of a maghemite particle described by a 
Heisenberg model, including exchange and dipolar interactions, 
using surface exchange and anisotropy constants different to those of the bulk. 
Their study was mainly focused on the thermal variation of the surface 
(for them consisting of a shell of constant thickness) and core magnetization, 
concluding that surface anisotropy is responsible for the non-saturation of the
magnetization at low temperatures. No attention was paid, however, to the magnetic 
properties under a magnetic field.

Other computer simulations studying finite-size and surface effects on ferro- 
and antiferromagnetic cubic lattices have also been published.
Trohidou et al. \cite{Trohidouprb90,Trohidoujap99} performed MC simulations of 
AF small spherical clusters. By using an Ising model on a cubic lattice 
\cite{Trohidouprb90}, they computed the thermal and magnetic field dependencies of 
the magnetization and structure factor, concluding that the particle behaved as a 
hollow magnetic shell. By means of a Heisenberg model \cite{Trohidoujap99}
with enhanced surface anisotropy, they studied the influence of different kinds 
of surface anisotropy on the magnetization reversal mechanisms 
and on the temperature dependence of the switching field. 
Dimitrov and Wysin \cite{Dimitrovprb94,Dimitrovprb95} studied the 
hysteresis phenomena of very small spherical and cubic FM fcc clusters of 
Heisenberg spins by solving the Landau-Lifshitz equations. 
They observed an increase of the coercivity 
with decreasing cluster size and steps 
in the loops due to the reversal of surface spins at different fields.
However they did not considered the finite temperature effects.


In order to contribute to elucidate the above mentioned experimental 
controversies and to further develop the previously published
numerical simulations, we present the results of a MC simulation of a single 
spherical particle which aim at clarifying what is the specific role 
of the finite size and surface on the magnetic properties of the particle, 
disregarding the interparticle interactions effects. 
In particular, we will study the magnetic properties under a magnetic 
field and at finite temperature, thus extending other simulation works.
In choosing the model, we have tried to capture the main features of
real particles with the minimum ingredients allowing to interpret
the results without any other blurring effects.

The rest of the article is organized as follows. In Sec. \ref{Sec2} we 
present the model of a maghemite particle upon which the MC simulations 
are based. In Sec. \ref{Sec3}, the study 
of the basic equilibrium magnitudes - energy, specific heat, and magnetization -
in absence of magnetic field is presented, comparing results for different particle 
sizes with those for periodic boundaries. Sec. \ref{Sec4} is devoted to the study of
magnetization processes under the presence of a magnetic field. 
The thermal dependence of hysteresis loops and coercive field are computed, 
and a detailed analysis of these quantities in terms of the surface and core 
contributions is performed. 
The effects of the introduction of different kinds of disorder 
on the magnetic properties are
presented in Sec. \ref{Sec5}, where we study both the deviation from ideal stoichiometry
by random removal of magnetic ions on the whole particle, as well as the introduction
of vacancies only at the surface of the particle. 
In Sec. \ref{Sec6}, we end up with a discussion of the obtained results and a 
presentation of the conclusions.

\section{Model}
\label{Sec2}
$\gamma$-Fe$_2$O$_3$, maghemite, is one of the most commonly studied  
nanoparticle compounds \cite{Kodamaprb99} presenting the above mentioned 
phenomenology.    
Maghemite is a ferrimagnetic spinel in which the magnetic Fe$^{3+}$ ions 
with spin $S= 5/2$ are disposed in two sublattices with different coordination  
with the O$^{2-}$ ions. Each unit cell (see Fig. \ref{Fe2O3_bw_fig}) has 8  
tetrahedric (T), 16 octahedric (O) sites, and one sixth  
of the O sites has randomly distributed vacancies to achieve neutrality charge.  
The T sublattice has larger coordination than O, thus, while the spins in the T  
sublattice have $N_{TT}=4$ nearest neighbours in T and $N_{TO}=12$ in O,  
the spins in the O sublattice have $N_{OO}=6$ nearest neighbours in O and  
$N_{OT}=6$ in T. In our model, the Fe$^{3+}$ magnetic ions are represented  
by Ising spins $S_i^{\alpha}$ distributed in two sublattices 
$\alpha=$ T, O of linear size 
$N$ unit cells, thus the total number of spin sites is ($24N^3$). 
The choice of Ising spins allows to reproduce a case with strong 
uniaxial anisotropy, while keeping computational efforts within 
reasonable limits. 
The possible existence of a spin-glass state at the surface of 
the particle should be better 
checked by using an Ising model than one with continuous spins, since in the former
frustration effects are enhanced \cite{Binder}.
Moreover, the Heisenberg version of the particle without disorder does not
show irreversibility in the hysteresis loops, whereas the Ising version does
\cite{Kodamaprb99,Soukoulisprl82}, being easier to observe independently the effects 
of disorder and finite size in the last case.

The spins interact via antiferromagnetic (AF) exchange interactions with the 
nearest neighbours on both sublattices and with an external magnetic field $H$,  
the corresponding Hamiltonian of the model being
\bea 
{\cal H}/k_{B}= 
-\sum_{\alpha,\beta=\,T,O}\sum_{i=1}^{N_\alpha}\sum_{n=1}^{N_{\alpha\beta}} 
          J_{\alpha\beta} S_i^{\alpha} S_{i+n}^{\beta}  \nonumber\\ 
          -h\sum_{\alpha= T,O}\sum_{i=1}^{N_\alpha} S_i^{\alpha}\ . 
\eea 
where we have defined the field in temperature units as 
\beq 
h=\frac{\mu H}{k_B} \ , 
\eeq 
being $S$ and $\mu$ the spin value and magnetic moment of the Fe$^{3+}$ ion,
respectively. 
Hereafter, $S_i=\pm 1$ and the maghemite values of the nearest neighbour 
exchange constants will be considered
\cite{Kodamaprb99,Kachkachieuj00}: $J_{TT}=-21$ K, $J_{OO}= -8.6$ K, $J_{TO}= -28.1$ K. 
Since the intersublattice interactions are stronger than those inside 
each sublattice, at low temperatures, there must be bulk ferrimagnetic 
order with spins in each sublattice ferromagnetically aligned and 
antiparallel intrasublattice alignment.
 
We have used periodic boundary conditions to simulate the bulk properties  
and free boundaries for a spherically shaped particle with $D$ unit cells  
in diameter, when studying finite-size effects. In the latter case, two  
different regions are distinguished in the particle: the surface formed by  
the outermost unit cells and an internal core of diameter $D_{Core}$ unit  
cells (see Fig. \ref{System2_fig}). 
The quantities measured after each MC step are the energy, specific heat,  
susceptibility and different magnetizations: sublattice magnetizations 
($M_O, M_T$),  
surface and core magnetization ($M_{Surf}, M_{Core}$), and total magnetization  
($M_{Total}$). Each of them have been normalized to the respective number of  
spins so that they can range from 1 to -1. In particular, $M_{Total}$ is $1$  
for ferromagnetic order, $0$ for a disordered system and $1/3$ for  
ferrimagnetic order of the O and T sublattices. 

The size of the studied particles ranges  
from $D= 3$ to $10$ corresponding to real particle diameters from $25$ to  
$83\ {\rm\AA}$ (see Table \ref{Table}). In this table, we have also included the 
number of surface and core spins $N_{Surf}, N_{Core}$, together with the
normalized magnetization values of a ferrimagnetic configuration $M_{Unc}$.
Note that due to the finite size of the particles, the ratio of T and O spins
produces $M_{Unc}$ values that, in general, do not coincide exactly with the 
$1/3$ value for perfect ferrimagnetic order in an infinite lattice. 
In order to make the measured magnetizations for different 
diameters comparable, we have normalized them to the corresponding
$M_{Unc}$ values given in Table \ref{Table}.

\section{Equilibrium Properties}
\label{Sec3}
 
\subsection{Energy and specific heat}

We start by studying the effect of free boundary conditions and  
finite-size effects on the equilibrium properties in zero magnetic field. 
The simulations have been performed using the standard Metropolis algorithm.  
Starting from a high enough temperature ($T= 200$ K) and an initially disordered state  
with spins randomly oriented, the system was cooled down at a constant  
temperature step $\delta T= -2$ K and, after discarding the first 1000 MC steps  
in order to allow the system to thermalize, the thermal averages of the  
thermodynamic quantities were computed at each temperature during a number  
of MC steps ranging from 10000 to 50000 depending on  
the system size.  
The starting configuration at each new temperature was the one obtained at the 
end of the averaging process at the previous temperature. Systems with periodic  
(PB) and free boundary (FB) conditions with spherical shape have been  
considered with sizes ranging from 3 to 14. 
 
In Fig. \ref{Fe2O3_E(T)_fig}, we compare the thermal dependence of the energy for 
spherical particles of different diameters $D$ with the corresponding 
results for a system of size $N=14$ and PB (lowermost curve, left triangles).  
A second order transition from paramagnetic to ferrimagnetic order  
signaled by a sharp peak at $T_c (D)$ in the specific heat  
(see the Inset in Fig. \ref{Fe2O3_E(T)_fig}) is clearly observed.  
Finite size effects on both the energy and the specific heat are very important  
even for $D$'s as large as 14 in the FB case, while for PB conditions they are 
negligible already for $N= 8$.  
The energy difference between the disordered and ferrimagnetic phases  
as well as the critical temperature $T_c (D)$ increases as $D$ is increased. 
This last quantity is strongly size dependent and approaches the infinite size 
limit ($T_c (\infty)= 126\pm 1$ K as evaluated for the $N= 14$ system with
PB conditions) as $D$ increases (see Fig. \ref{Fe2O3_Tc(D)_fig}, 
in which the variation of the peak in the specific heat with $1/D$ has been 
plotted). $T_c (D)$ can be fitted to the scaling law  
\beq 
\frac{T_{c}(\infty)-T_c (D)}{T_{c}(\infty)} = \left(\frac{D}{D_0}\right) ^{-1/\nu} 
\label{TcEq}
\eeq 
as predicted by finite-size scaling theory \cite{Landauprb76,Barber} with 
$D_0 = 1.86\pm 0.03$ a microscopic length scale (in this case, it is roughly
twice the cell parameter),  
and a critical exponent $\nu=0.49\pm 0.03$, which seems to indicate a mean 
field behaviour \cite{Stanley}. This result can be 
ascribed to the high coordination of the O and T sublattices.
The fitted curve is drawn in Fig. \ref{Fe2O3_Tc(D)_fig} where deviations from 
scaling are appreciable for the smallest diameters for which corrections to 
the finite-size scaling of Eq. \ref{TcEq} may be important \cite{Landauprb76}.  
 
\subsection{Magnetization}

To study the effects of a free surface and of finite size on the magnetization 
of the particles, we compare in  Fig. \ref{Fe2O3_M(T)H0_fig} the 
results for four particle diameters ($D= 3, 4, 6, 8$, open circles) with that 
corresponding to a $N= 14$ system with PB (representing the behaviour  
of the bulk). In this figure, we have distinguished the surface (dashed lines) 
and core (dot-dashed lines) contributions to the total magnetization (symbols). 
The results have been recorded during the same cooling procedure used to  
obtain the energy.   
The main feature observed is the reduction of the total magnetization  
$M_{Total}$ with respect to the PB case (continuous line) due to the lower  
coordination of the spins at the surface, which hinders ferrimagnetic order  
at finite temperatures. 
Fig. \ref{Fe2O3_M(T)H0_fig} clearly shows the roles played by the surface and 
the core in establishing the magnetic order. On one hand, 
independently of the particle size, the core (dot-dashed lines) tends to a  
perfect ferrimagnetic order at low $T$ (marked by $M= 1/3$), progressively  
departing from the bulk behaviour as $T$ approaches $T_c$,  
this finite-size effect being more important as the particle size decreases.
However, the surface magnetization does not attain perfect ferrimagnetic order
at $T=0$ even for $D=8$ due to the reduced coordination of the spins. 
For this reason, a rapid thermal demagnetization is observed which 
significantly departs $M_{Surf}$ from the bulk behaviour. 
 
It is worthwhile to note that for all the diameters studied there is a 
temperature range in which this demagnetization process is linear, this range
being wider as the particle size decreases. 
In this linear regime, the particle demagnetization becomes dominated by the 
surface effects, being the core and surface behaviours strongly correlated. 
Linear demagnetization is indicative of the effective 3D-2D dimensional 
reduction of the surface shell and has previously been observed in thin film 
systems \cite{Martinezjpcm92,Parkprl98} and in simulations 
of rough FM surfaces \cite{Zhaoprb00}. 
$M_{Total}$ is always strongly dominated by the surface contribution, 
progressively tending to the bulk behaviour as the particle size is increased.

\section{Hysteresis Loops} 
\label{Sec4}
 
In Fig. \ref{Fe2O3_HIDnv0T_fig}, we show the hysteresis loops of particles with diameters  
$D=3, 6$ for different temperatures. The loops have been computed by starting  
from a demagnetized state at $h= 0$ and increasing the magnetic field in  
constant steps, $\delta h= 1$ K, during which the magnetization was averaged  
over $\approx 3000$ MC steps after thermalization.  
The results shown have been averaged for several independent runs starting  
with different random seeds.  
 
First of all, let us note that the saturation field and  
the high field susceptibility increase as the particle size is reduced,  
since this quantities are mainly associated to the  
progressive alignment of the surface spins towards the field direction.  
Thus, the loops of the smallest particles resemble those found in 
ferrimagnetic nanoparticles \cite{Kodamaprl96,Kodamaprl97,Kodamaprb99}
and other bulk systems with disorder \cite{Binder,Maletta81}, 
increasing their squaredness 
(associated to a uniform reversal mechanism of $M$) with the size.   
In fact, by plotting separatedly the contributions of the core and the surface 
to the total magnetization (see Fig. \ref{Fe2O3_HIDnv0SC_fig}, dashed lines), 
we see that the loop of the core is almost perfectly squared independently 
of temperature and particle size, indicating a uniform reversal of its 
magnetization with a well-defined ferrimagnetic moment. 
Instead, the loop of the surface reveals a progressive reversal of $M$, 
which is a typical feature associated to disordered or frustrated systems 
\cite{Binder,Maletta81}.
Nonetheless, for a wide range of temperatures and particle
sizes, it is the reversal of the surface spins which triggers the reversal 
of the core. This is indicated by the fact that the coercive field of 
the core is slightly higher but very similar to the one of the surface. 
 
Since for all the studied particle sizes the $h_c(T)$ 
curves show a complex behaviour mainly related to the frustration of the 
antiferromagnetic intra and intersublattice exchange interactions,
we start by studying the case of a ferromagnet with no frustration. 
In Fig. \ref{Fe2O3_HCoeDnv0_fig}a, the $h_c(T)$ dependence for a system 
with $N=8$ and with
the same lattice structure as maghemite but equal FM interactions 
$J_{\alpha \beta}=J$ and PB conditions is shown.
The $h_c(T)$ dependence is now a monotonously decreasing curve 
with no inflection point, 
which at high enough temperatures ($T/J\gtrsim 1$) can be fitted to a power law of the kind
\beq 
h_{c}(T)= h_{c}(0)[1-\left(T/T_{c}\right)^{1/\alpha}] \ , 
\label{Hcoe} 
\eeq 
with $\alpha= 2.26\pm 0.03$; close but different to what would be obtained by a model
of uniform reversal such as Stoner-Wohlfarth \cite{Stoner48} ($\alpha=2$).
Even in this simple case, for which $M$ reverses uniformly, 
the thermal variation of $h_c(T)$ cannot be only ascribed to the
thermal activation of a constant magnetization vector over an energy barrier 
landscape, since actually $M$ is of course temperature dependent.
Therefore, the reversal mechanism cannot be inferred from the $\alpha$ value
obtained from a fit to Eq. \ref{Hcoe} in any range of temperatures for which
$M$ significantly varies with $T$.

The thermal dependence of $h_c$ for the maghemite particles with AF interactions is shown in 
Fig. \ref{Fe2O3_HCoeDnv0_fig}b. 
Both for the PB and spherical cases, the $h_c(T)$ curves are qualitatively different from
the FM case: they have opposite curvature and two regimes of thermal
variation.

Let us start by analyzing the PB case. At high $T$ ($T\gtrsim 20$ K), $h_c(T)$ can be 
fitted to the power law of Eq. \ref{Hcoe} with $\alpha = 0.94\pm 0.02, 
h_c(0)=134\pm 2$ K. 
Values of $\alpha$ close to 1 have been deduced in the past for some 
models of domain wall motion \cite{Gaunt}. 
At low $T$, a different regime is entered but tending to the same $h_c(0)= 134.2$ K. 
This change in behaviour is associated to the wandering of the system through  
metastable states with $M_{Total}\simeq 0$, which are induced by the frustration 
among AF interactions. Consequently, when lowering $T$, the hysteresis loops 
become step-like around $h_c$ as shown in Fig. \ref{Fe2O3_HIPBT_0_fig}
(similar features are observed in related studies 
\cite{Dimitrovprb94,Dimitrovprb95,Vittalaprb97}).  
The jumps at $T= 0$ are located at $h= 117, 134.2, 151.4$ K, the values at which 
the magnetic field energy is enough to invert one O spin having $0, 1, 2$
O nearest neighbours inverted, respectively \cite{Vittalaprb97}. 
While the O sublattice does not reverse uniformly, 
the T sublattice instead reverses as a whole after the reversal of O, at $h= 151.4$ K. 
When $T$ is increased from 0, the steps are rounded by the progressive population 
of states with greater degree of configurational disorder and less metastability, 
finally giving rise to the suppression of the steps for $T$ around
$12$ K, when $h_c\lesssim 117$ K and the high $T$ regime of $h_c(T)$ is entered.

The general $h_c(T)$ behaviour for spherical particles with FB strongly depends on the 
particle size. For $D= 3, 6$ and $T\gtrsim 20$ K, the $h_c$ decay is similar to that 
for PB, but, at any given $T$, being smaller than for PB and as the size of the 
particle is decreased. At these temperatures, $h_c(T)$ is dominated by the surface,
which nucleates the reversal of the magnetization, as indicated by the proximity
between the surface and core $h_c$ (see Fig. \ref{Fe2O3_HIDnv0SC_fig}). 
However, when lowering $T$ below $20$ K, $M_{Surf}$ and $M_{Core}$ tend to be equal, 
the surface becomes less efficient as nucleation center for spin reversal,  
and $h_c$ becomes dominated by the core ($h_c^{Surf}<h_c^{Core}$ for any particle size
, see Fig. \ref{Fe2O3_HIDnv0SC_fig}).
This is the cause of the rapid increase of $h_c$ towards the PB values for the
$D=6$ curve (see Fig. \ref{Fe2O3_HCoeDnv0_fig}b). For $D=3$, instead, $h_c$
saturates when lowering $T$ due to the smaller ratio of core to surface spins,
which actually hinders the prevalence of the core.

Finally, it is worth noticing that, independently of the size of the particles
with FB, the $h_c$ values are always smaller than that for PB,
since the existence of spins with less coordination at the surface favours 
the formation of reversed nuclei of spins acting as a seed for the 
reversal process, which is not the case of PB, where all equivalent spins  
have the same coordination. Therefore, the $h_c$ values for PB are only recovered
at low T in the limit of large particle size, at difference with other extensive 
magnitudes such as the energy or the magnetization, for which we have checked 
that finite-size scaling is accomplished.

\section{Effects of disorder} 
\label{Sec5}

In real particles, disorder and imperfections are present   
departing the system from perfect stoichiometry and distort the position 
of the atoms on the lattice, being their effect more important at the 
surface \cite{Moralesjpcm97}. 
There are several ways to implement this disorder on the model. 
The simplest way to simulate the deviation of the O and T sublattice atoms 
from ideal stoichiometry is by random removal of magnetic ions on the  
O/T sublattices.   

\subsection{Disorder on the lattice}

Up to the moment, the existence of vacancies in the O sublattice in real 
maghemite structure has not been considered. It is important to note
that, in this system, intra and intersublattice magnetic interactions are 
antiferromagnetic. Consequently, inclusion of vacancies in one of the
sublattices may destabilize the FM parallel alignment of the 
other one, resulting in a system with a great degree of magnetic disorder.
In particular, this effect will be much stronger when vacancies are 
introduced in the O sublattice, since $N_{OT}$ is greater than $N_{TO}$. 
To show the effect of these kind
of disorder, we have simulated the hysteresis loops for different vacancy
concentrations $\rho_v$ on the O sublattice at two cooling fields 
$h_{FC}= 20, 100$ K. As can be seen in Fig. \ref{Fe2O3_HIDnvT20_fig},
the introduction of a low concentration of vacancies ($\rho_{v}=1/6$ as in
the real material) results in a reduction of the magnetization and 
increases the high field susceptibility without any substantial change 
in the general shape of the loops. 
However, if $\rho_{v}$ is increased 
beyond the actual value, the loops progressively closes, loosing  
squaredness and progressively resembling those for a disordered 
\cite{Binder,Maletta81} system, 
with high values of the high field susceptibilities and much lower coercivity. 

\subsection{Surface disorder}

In what follows, we will study the effects of the disorder at the surface
of the particle, considering a $\rho_v = 1/6$ vacancy density on the O 
sublattice. Since the surface of the particles is not an ideal sphere, 
the outermost unit cells may have an increased number of vacancies on both  
sublattices with respect to those present in the core. Reduced coordination
at the surface may also change the number of links between the surface atoms. 
We will be denote by $\rho_{sv}$ the concentration of surface vacancies
in the outermost primitive cells.

\subsubsection{Field Coolings}

The magnetic ordering of the system can be characterized by studying the 
behaviour of the equilibrium magnetization in a magnetic field. These curves 
have been obtained by the same cooling procedure used in the magnetization  
simulations at zero field with $\delta T= -2$ K in presence of different 
cooling fields $h_{FC}$. 
Several such curves are shown in Figs.  
\ref{Fe2O3_FCDsvD3_fig}, \ref{Fe2O3_FCDsvD6_fig}, 
in which the surface (continuous lines) and the core 
(dashed lines) contributions to the total magnetization (open symbols) 
have been distinguished.
Let us first analize the case with no surface disorder ($\rho_{sv}= 0$). 
The curves at different cooling fields do not collapse to the perfect ferrimagnetic
order value at low $T$ (i.e. $M_{Total}= 1/4$ for $\rho_{sv}$=1/6), 
reaching higher values of 
the magnetization the higher $h_{FC}$, being this effect greater as
the particle size is reduced (compare Fig. \ref{Fe2O3_FCDsvD3_fig}a and 
Fig. \ref{Fe2O3_FCDsvD6_fig}a). 
This is in contrast with the results for PB (not shown), for which the system reaches 
perfect ferrimagnetic order at low $T$, even at fields higher than $100$ K, 
evidencing that the main effect of the surface is the breaking of 
ferrimagnetic correlations within the particle. As a consequence, at a given 
temperature, the FM order induced by a magnetic field increases 
when decreasing $D$.

By separatedly analyzing in detail the behaviour of the surface and core 
contributions to the total magnetization, deeper understanding of 
finite-size effects can be gained.  
As in the case of $h=0$, the total magnetization for small particles is  
completely dominated by the surface contribution (continuous lines  
in Fig. \ref{Fe2O3_FCDsvD3_fig}, \ref{Fe2O3_FCDsvD6_fig}) and this 
is the reason why the ferrimagnetic order is less perfect at these 
small sizes and the magnetic field can easily  magnetize the system.  
However, the behaviour of the core of the smallest particles is still  
very similar to that of the case with PB, although its contribution  
to $M_{Total}$ is very small.  
At low fields, the surface is always in a more disordered state than the core:  
its magnetization lies below $M_{Total}$ at temperatures for which the thermal  
energy dominates the Zeeman energy of the field (see the continuous lines  
in Fig. \ref{Fe2O3_FCDsvD3_fig}a, \ref{Fe2O3_FCDsvD6_fig}a).  
In this regime, the total magnetization closely follows 
that of the surface (see the curves in Figs. \ref{Fe2O3_FCDsvD3_fig}a and 
\ref{Fe2O3_FCDsvD6_fig}a for $h_{FC}= 20$ K)
for the two simulated sizes.
In contrast, a high field is able to magnetize the surface easier than the  
core due to the fact that the broken links at the surface worsen the  
ferrimagnetic order, while the core spins align towards the field  
direction in a more coherent way. Only for the biggest particles the 
surface contribution departs from the $M_{Total}$ indicating the 
increasing contribution of the core (see the curves in Figs. 
\ref{Fe2O3_FCDsvD3_fig}a and \ref{Fe2O3_FCDsvD6_fig}a for $h_{FC}= 20$ K). 
Note also, that in this high $h_{FC}$ regime,  
a maximum appears which is due to the competition between the FM 
alignment induced by the field and the spontaneous ferrimagnetic order 
(as the temperature is reduced the strength of the field is not enough 
as to reverse the spins into the field direction).  

The introduction of vacancies does not change the low field behaviour of 
the total magnetization, which is still dominated by the surface both for
$D= 3, 6$, although the smallest particles are easily magnetized by the field.
However, at high fields, $M_{Total}$ is lower than $M_{Surf}$, the surface 
progressively decouples from $M_{total}$ with the introduction 
of vacancies in the surface, being this effect more remarkable for the
biggest particle.
With respect to the core, at difference with the non disordered case
($\rho_v = \rho_{sv}=0$), 
the low temperature plateau of $M_{Core}$ tends to a higher value than 
that for perfect ferrimagnetic order, since the main effect of the disorder 
is to break ferrimagnetic correlations in the core; increasing the 
ferromagnetic order induced by the field. This is reflected in a progressive
departure of the high and low field $M_{Core}$ curves with increasing 
disorder (see the dashed lines in the sequence b-d of Figs. 
\ref{Fe2O3_FCDsvD3_fig}, \ref{Fe2O3_FCDsvD6_fig}). The maximum appearing 
at high $h_{\rm FC}$ is only slightly affected by disorder, shifting 
to lower temperatures and eventually disappearing for $D=3$ and $\rho_{sv}=0.5$.

\subsubsection{Hysteresis loops}

Hysteresis loops with surface disorder are given in Fig. 
\ref{Fe2O3_HIDsvT20_fig} for two particle diameters.
The introduction of surface vacancies facilitates the magnetization 
reversal by progressive rotation, producing a 
rounding of the hysteresis loops when approaching $h_c$, in the same way 
that occurs when particle size is reduced.
The same fact explains the increase of the high field susceptibility, 
since the vacancies act as nucleation centers of FM domains 
at the surface, which, from there on, extend the FM correlations 
to the inner shells of spins. 
Moreover, a considerable decrease of $h_c$ is observed. All these facts
yield to a progressive elongation of the loops, giving loop shapes 
resembling those of disordered systems \cite{Binder,Maletta81}. 
Fig. \ref{Fe2O3_HIDsvSC_fig}, where the surface and core contributions are
shown separatedly, clearly evidences that the increase of FM 
correlations at the surface, facilitated by the vacancies, induce FM 
order in the core. 
That is to say, $M_{Core}$ follows the evolution of $M_{Surf}$ at moderate 
fields above $h_c$, in contrast with the case with no surface 
vacancies (see Fig. \ref{Fe2O3_HIDnv0SC_fig}) where the core keeps 
the ferrimagnetic order for the same field range.   

\section{Discussion and conclusions}
\label{Sec6} 

We have presented a simple model of a maghemite nanoparticle with the minimal
ingredients necessary to faithfully reproduce the magnetic structure of the real material.
The model has proven successful in reproducing several key features present in ferrimagnetic nanoparticle systems: (1) the reduction of $T_c$, spontaneous
magnetization $M_{Total}$, and coercive field $h_c$, for small sizes, as $D$ decreases; 
(2) the increase, with the reduction of the particle size and with the 
increase of surface disorder, of the differential susceptibility and 
the elongation of the hysteresis loops in resemblance with those of 
frustrated systems; and (3) the existence of a surface layer with 
higher magnetic disorder than the core. Let us further comment these 
points in deeper detail.

First of all, we find that $T_c(D)$ follows conventional finite-size scaling, 
discarding any important surface effect on this quantity. Similar 
finite-size effects have been found in fine particles \cite{Tangprl91} 
of MnFe$_2$O$_4$, but with a surprising increase of $T_c(D)$ as $D$ decreases.
However, the spontaneous magnetization $M_{Total}$, at any temperature, follows
a quasi-linear behaviour with $1/D$, see Fig. \ref{Fe2O3M(D)_fig}, indicating 
that the reduction of $M_{Total}$ is simply proportional to the ratio of 
surface to core spins, so it is mainly a surface effect. Similar experimental
behaviour has been found in $\gamma$-Fe$_2$O$_3$ \cite{Hanjm94} and 
the above mentioned system \cite{Tangprl91}.

The $h_c(T)$ thermal decay for the spherical particles is in qualitative
agreement with the experimental results for maghemite particles 
of sizes 9-10 nm shown in Fig. 4 of Ref. \onlinecite{Martinezprl98}, 
taking into account that in real samples there are additional contributions
coming from the blocking process associated to the particle size 
distribution ($h_c$ drops to zero above the blocking temperature).
In both cases, the curvature of the $h_c(T)$ curve is similar,
suggesting a non-uniform reversal of the magnetization, 
a point that is also confirmed by the shape of the hysteresis 
loops around $h_c$. 
However, our model for spherical particles gives reduced coercivities 
with respect to the bulk (represented by the PB case). 
A fact that is in contrast with the enhancement observed experimentally \cite{Kodamaprl97,Martinezprl98,Batlleprb97},
and indicating that finite-size effect cannot cause it.  
Increased anisotropy at the surface may be the responsible for it.
In any case, the model qualitatively reproduces the $h_c$ reduction
with $D$ for small sizes [see Fig. 1 in Ref. \onlinecite{Kodamaprl97}], 
which may be indeed a finite-size effect.

The $M(T)$ and $M(h)$ dependencies obtained in our simulation lead to 
the conclusion that, in spherical particles, there is a surface layer 
with much higher degree of magnetic disorder than the core, 
which is the Ising version of the random  
canting of surface spins occurring in several fine particle with spinel
structure \cite{Coeyprl71,Morrishjap81,Kodamaprl96,Linjm95,Jiangjpcm99}. 
As opposite with the suggestion given by some authors 
\cite{Kodamaprl97,Martinezprl98} that below a certain freezing
temperature the surface layer enters a spin-glass-like state, our model
does not give any indication of this phenomenology at any of the studied 
sizes and temperatures. 
Furthermore, the surface layer, by partially breaking the ferrimagnetic 
correlations, diminishes the zero-field $M_{Total}$ but, 
at the same time, enhances $M_{Total}$ at moderate fields. 
Although the surface is easily thermally demagnetized and easily magnetized 
by the field than the core, it does not behave as a dead layer, since, 
at any $T$, it is magnetically coupled to the core. 
All these facts put forward that the surface has higher magnetic response 
than the core, excluding a spin-glass freezing. 
Moreover, we do not observe irreversibilities between field and 
zero-field cooled magnetization curves, which is a key signature that in 
the scope of our model, neither finite-size or surface effects, 
nor the inclusion of surface vacancies are enough to account for 
the postulated spin-glass-like state. 

Finally, let us mention that our model does not reproduce the 
experimentally observed shift of the hysteresis loops under field
cooling adduced as a prove of the existence of the spin-glass-like 
state at the surface \cite{Kodamaprl97,Kodamaprb99,Martinezprl98}.
Only when $h_{FC}$ smaller than irreversibility fields $h_{irr}$
are used in the numerical experiment, hysteresis loops that are
apparently shifted are obtained, which in fact are minor loops.
In any case, the absence of this phenomenology is in agreement with the
non-observation of a spin-glass-like state at the surface,
indicating that other $\sl {ad\ hoc}$ ingredients must be included
in the model. For instance, enhanced surface anisotropy or exchange 
constants at the surface different than at the bulk, as is the case in
exchange coupled multilayers \cite{Nowakprl00,Stampsjpd00}.

Current work is under progress to elucidate the possible influence 
of these new ingredients and of interparticle interactions.
 
\section*{Acknowledgements}
The authors are indebted to Prof. F\`elix Ritort for close collaboration in the
initial steps of this study and critical reading of the manuscript.
We acknowledge CESCA and CEPBA under coordination of 
C$^4$ for the computer facilities. This work has been supported by 
SEEUID through project MAT2000-0858 and CIRIT under project 2000SGR00025.

\newpage
\begin{centering} 
\begin{table*} 
\setlength{\tabcolsep}{8.5pt}
\caption{Characteristic parameters of some of the spherical particles simulated: 
particle diameter $D$ in units of the lattice constant $a$,  
diameter of the corresponding real $D_{Real}$, number of  
total spins $N_{Total}$, number of spins at the surface and in the core  
$N_{Surf}, N_{Core}$, and magnetization of the noncompensated spins  
$M_{Unc}=(N_O - N_T)/N_{Total}$. 
The data are for particles with no vacancies in the O sublattice.  
}  
\begin{tabular}{ccccccc} 
        D	& $\rm D_{ Real}$ (\AA)& $\rm N_{Surf}$	& $\rm N_{Core}$ 	& $\rm M_{Unc}^{Surf}$ &  $\rm M_{Unc}^{Core}$	& $\rm M_{Unc}$ \\ \hline  
        3   & 25          & 330  (95\%) & 17        & 0.285  & 0.412 & 0.291 \\  
        4   & 33          & 724  (87\%) & 111       & 0.337  & 0.369 & 0.341 \\  
        5   & 41          & 1246 (78\%) & 347       & 0.355  & 0.291 & 0.341 \\  
        6	& 50		  & 1860 (69\%)	& 841		& 0.350  & 0.332 & 0.344 \\  
        8   & 66          & 3748 (58\%) & 2731      & 0.345  & 0.330 & 0.338 \\    
        10	& 83	 	  & 6485 (48\%)	& 12617		& 0.329  & 0.337 & 0.333 \\    
    \end{tabular}  
\label{Table} 
\end{table*}  
\end{centering} 
\bfig[htbp] 
\centering 
\leavevmode  
\epsfxsize=8 cm  
\epsfbox{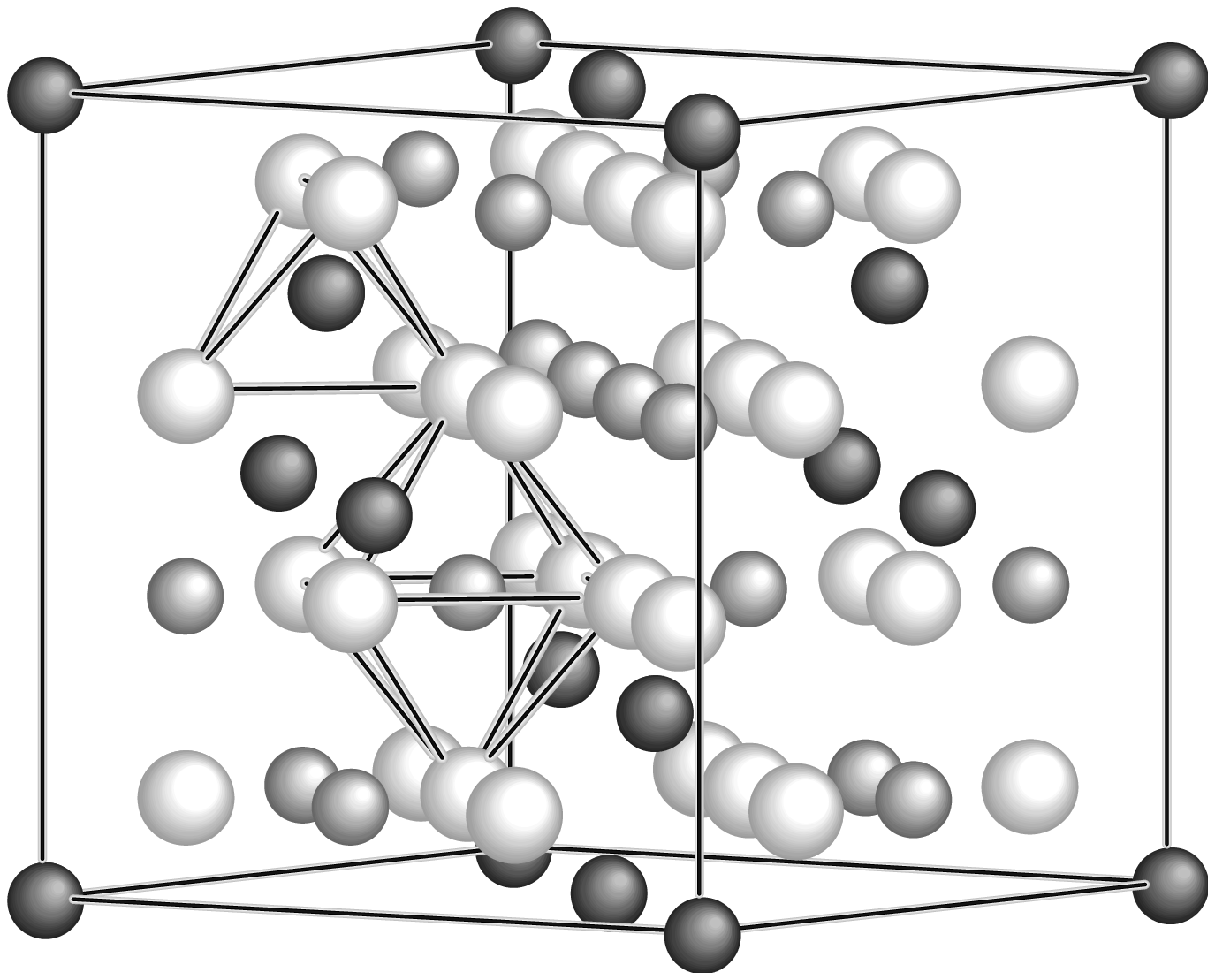}  
\caption{Unit cell of maghemite. The magnetic Fe$^{3+}$ ions occupying the two 
sublattices, in different coordination with the O$^{2-}$ ions (white colour), are  
coloured in black (T sublattice, tetrahedric coordination) and in grey  
(O sublattice, octahedric coordination). 
}  
\label{Fe2O3_bw_fig} 
\efig 
\bfig[htbp] 
\centering 
\leavevmode  
\epsfxsize=8 cm  
\epsfbox{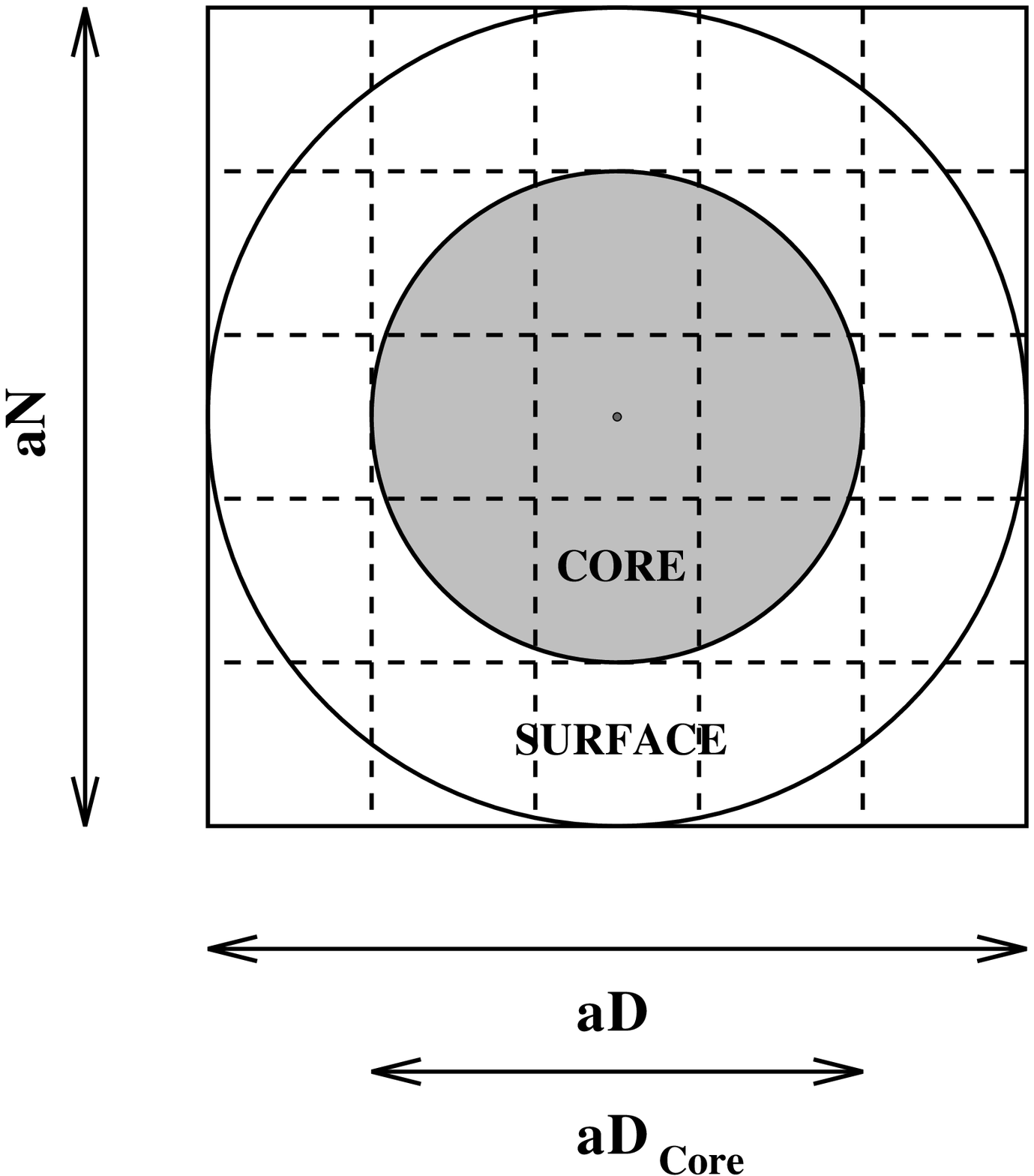}  
\caption{Schematic drawing of the spherical particles simulated in this study,  
showing the basic geometric parameters. The unit cells are indicated by the 
dashed grid, being the cell parameter $a$, and $N$ the number of unit cells
along each axis.
}  
\label{System2_fig} 
\efig 
\bfigu[htbp] 
\centering 
\leavevmode  
\epsfxsize=13 cm  
\epsfbox{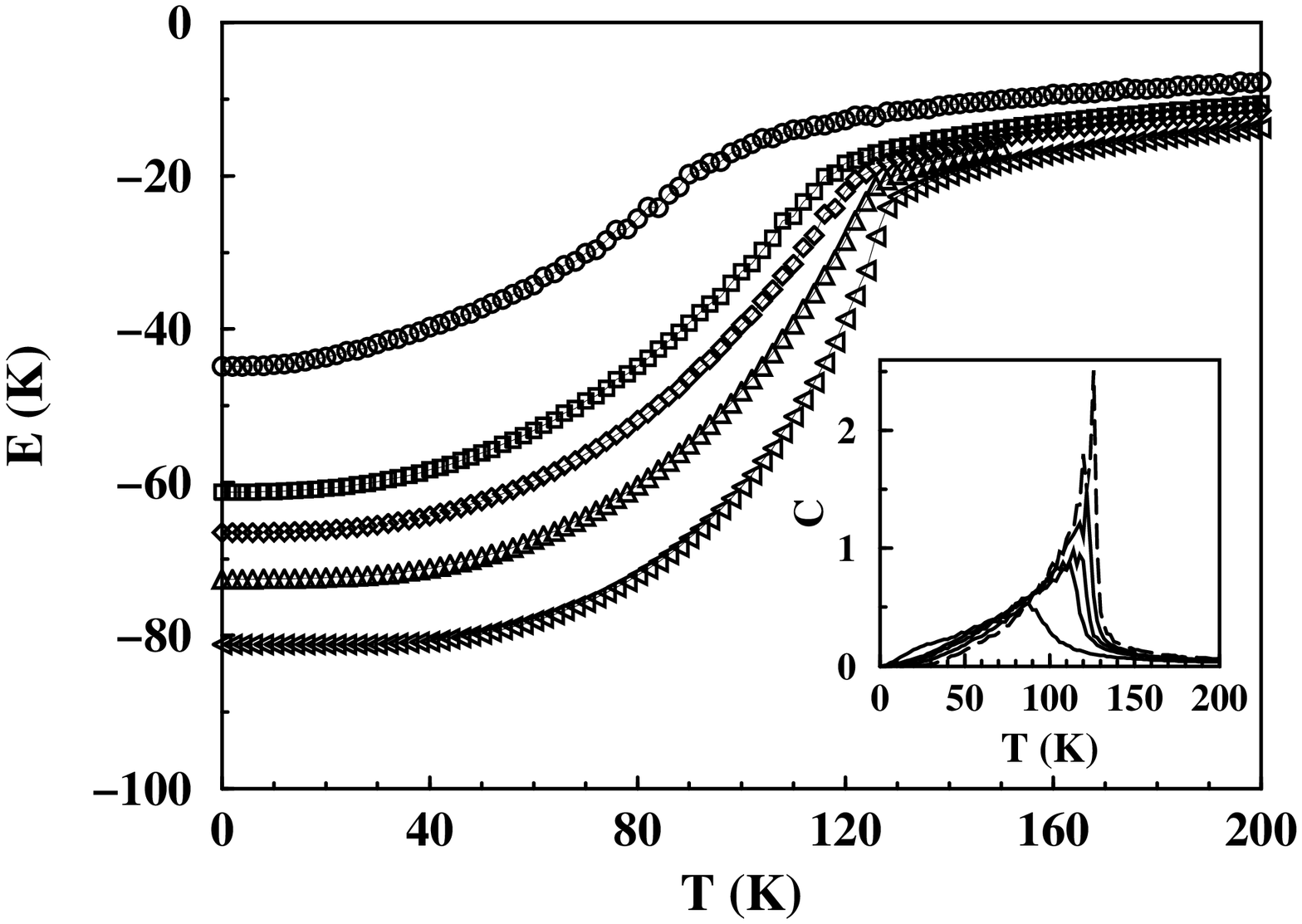}  
\caption{Thermal dependence of the energy for different diameters $D= 3, 6, 8,  
14$ (from the uppermost curve) and periodic boundary conditions  
$N= 14$ (lowermost curve). Inset: Thermal dependence of the specific heat for  
the same cases (the periodic boundary case is drawn with a dashed line). 
}  
\label{Fe2O3_E(T)_fig} 
\efigu 
\bfigu[htbp] 
\centering 
\epsfxsize=13 cm  
\leavevmode  
\epsfbox{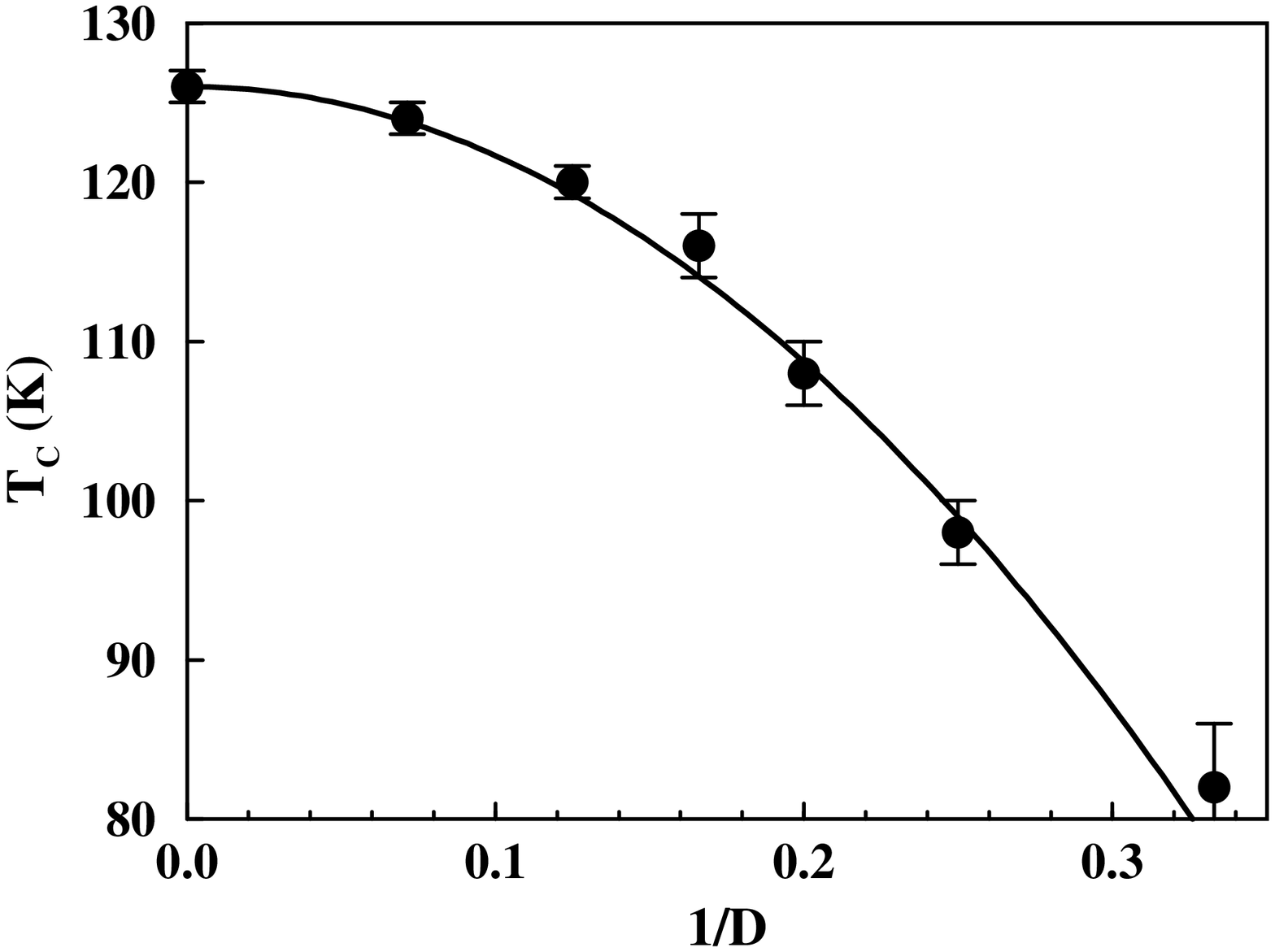}  
\caption{Particle size dependence of the transition temperature $T_c$ from 
paramagnetic to ferrimagnetic phases for spherical particles with FB. 
The displayed values have been obtained from the maximum in the  
specific heat. The continuous line is a fit to Eq. 3.
}  
\label{Fe2O3_Tc(D)_fig} 
\efigu 
\bfigu[htbp] 
\centering 
\epsfxsize=13.4 cm  
\leavevmode  
\epsfbox{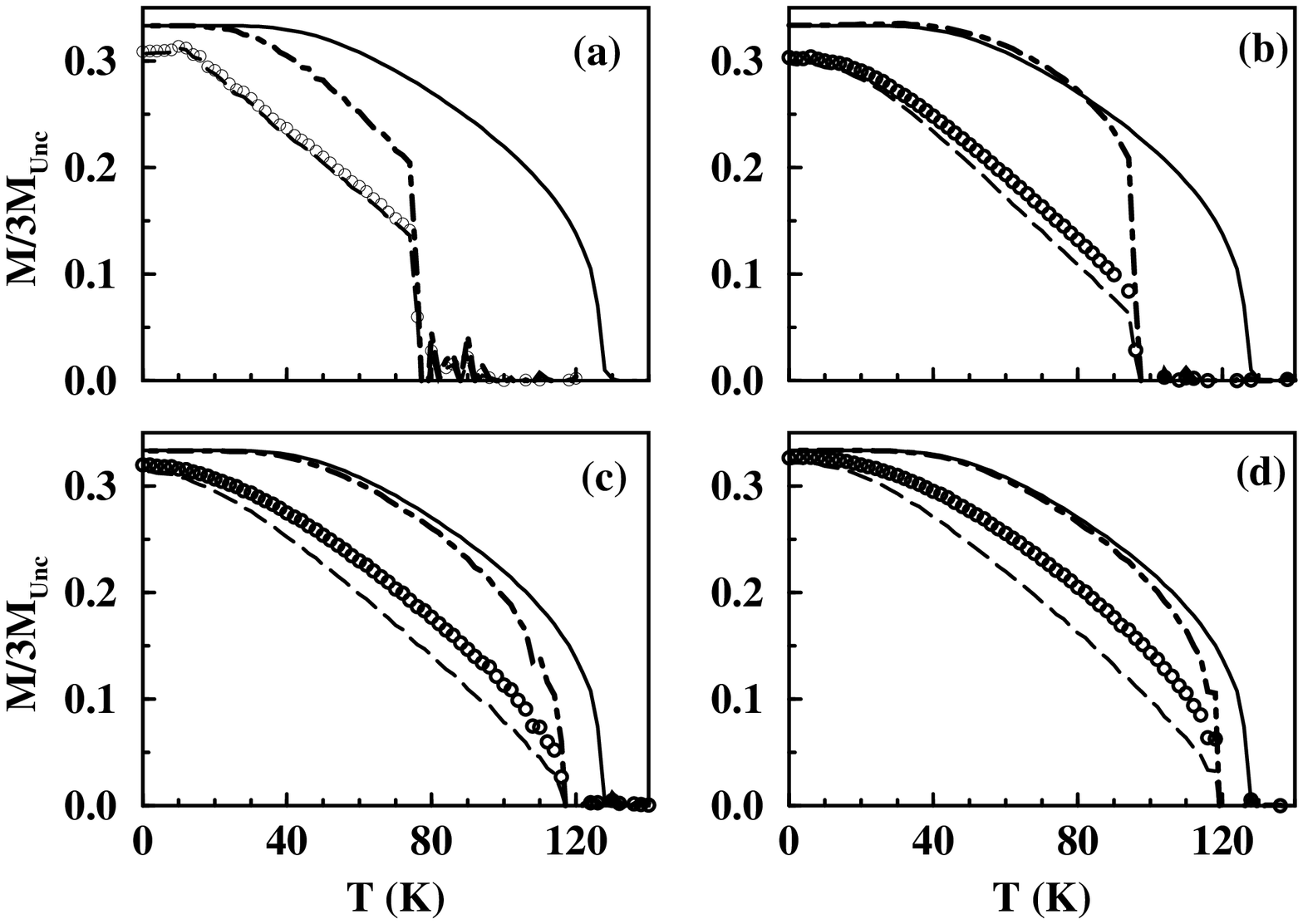}  
\caption{Thermal dependence of the magnetization $M$ obtained by progressive  
cooling from high $T$ at a constant rate, $\delta T= -2$ K, and starting from a 
random configuration of spins. The results for four particle diameters  
are shown: $D= 3$ (a), $D= 4$ (b), $D= 6$ (c), and $D= 8$ (d).  
The contributions of the surface  
(dashed line) and core spins (dot-dashed line) have been distinguished  
from the total magnetization (circles).  
The results for PB conditions, in a system of linear size  
$N= 14$, have also been included for comparison (continuous line).  
}  
\label{Fe2O3_M(T)H0_fig} 
\efigu 
\bfigu[htbp] 
\centering 
\epsfxsize=13 cm  
\leavevmode  
\epsfbox{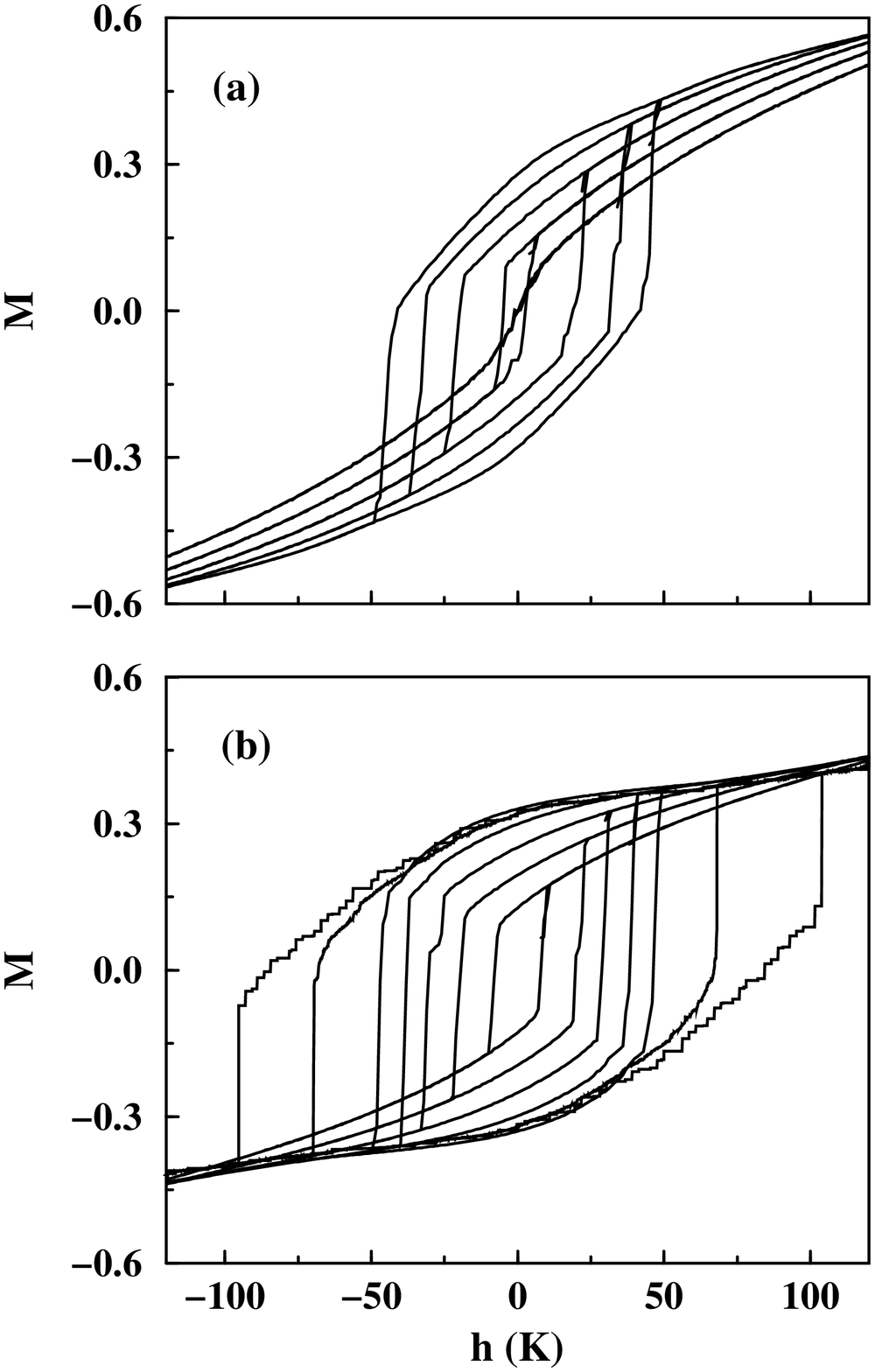}  
\caption{Temperature dependence of the hysteresis loops for particles of diameter 
$D=3$ (a), $D=6$ (b). The temperatures starting from the outermost loop  
are $T= 0, 5, 20, 40, 60, 80, 100$ K, except for case $D=3$ in which $T= 0, 5$ K
curves are omitted for clarity.}  
\label{Fe2O3_HIDnv0T_fig} 
\efigu 
\bfigu[htbp] 
\centering 
\epsfxsize=13 cm  
\leavevmode  
\epsfbox{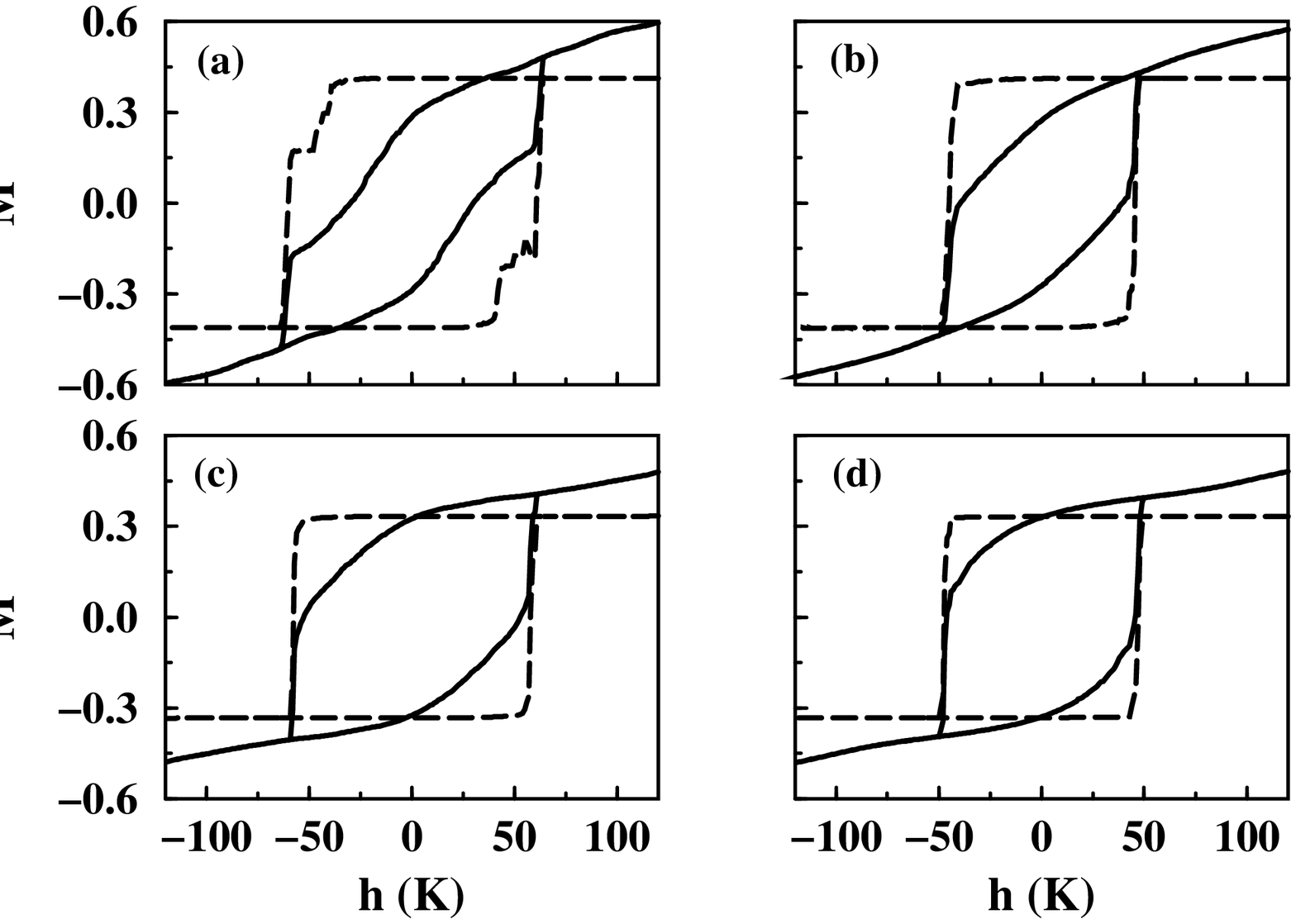}  
\caption{Surface (continuous line) and core (dashed line) contributions to the  
hysteresis loops for particles of diameters $D= 3$, $T= 10$ K (a); 
$D= 3$, $T=20$ K (b);
$D= 6$, $T= 10$ K (c); $D= 6$, $T=20$ K (d).}  
\label{Fe2O3_HIDnv0SC_fig} 
\efigu 
\bfigu[htbp] 
\centering 
\epsfxsize=13 cm  
\leavevmode  
\epsfbox{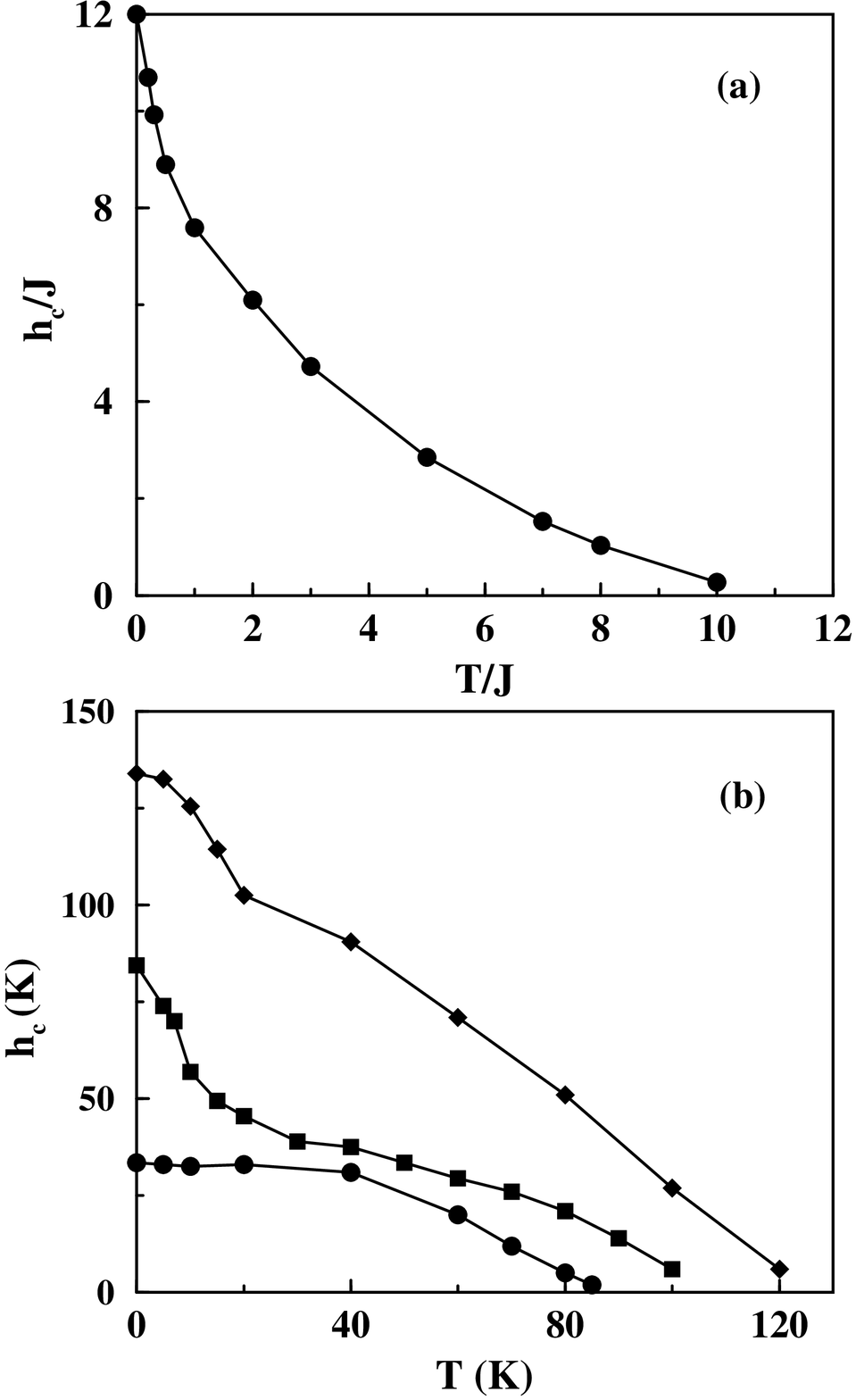}  
\caption{ 
(a) Temperature dependence of the coercive field $h_c$ for a system with
the same structure as maghemite but ferromagnetic interactions 
($J_{\alpha\beta}=J$) and PB conditions and $N= 8$; 
(b) Temperature dependence of the coercive field $h_c$
for the real AF values of the exchange constants for maghemite for
the case of FB spherical particles of diameters 
$D= 3$ (circles), $D= 6$ (squares), and for a system of linear size
$N=8$ with PB conditions (diamonds).}  
\label{Fe2O3_HCoeDnv0_fig} 
\efigu 
\bfigu[htbp] 
\centering 
\epsfxsize=13 cm  
\leavevmode  
\epsfbox{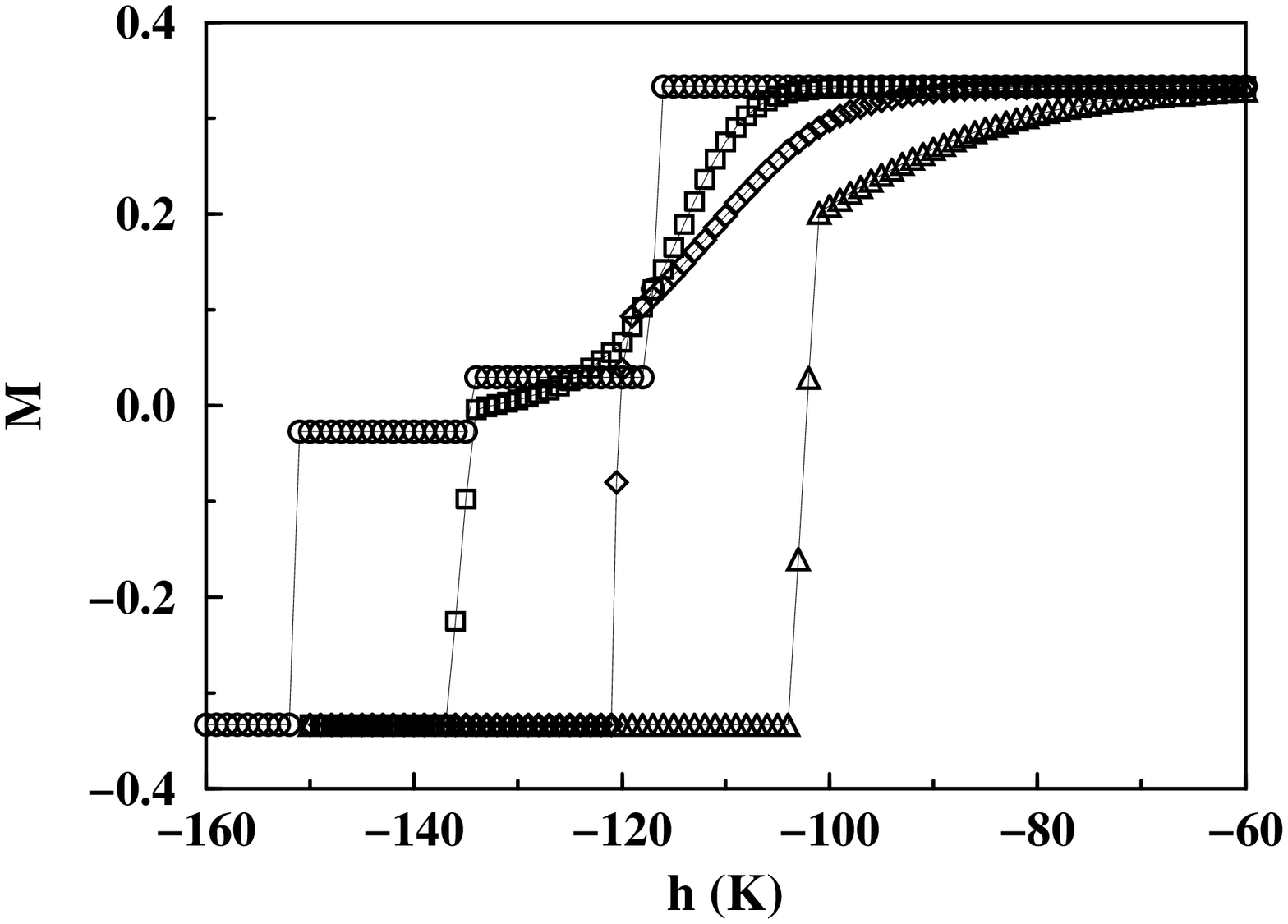}  
\caption{Detail of low temperature hysteresis loops for PB conditions
around the coercive field $h_c$. The corresponding temperatures are
$T=$ 0 (circles), 5 K (squares), 10 K (diamonds), 20 K (triangles).
}  
\label{Fe2O3_HIPBT_0_fig} 
\efigu 
\bfigu[htbp] 
\centering 
\epsfxsize=13 cm  
\leavevmode  
\epsfbox{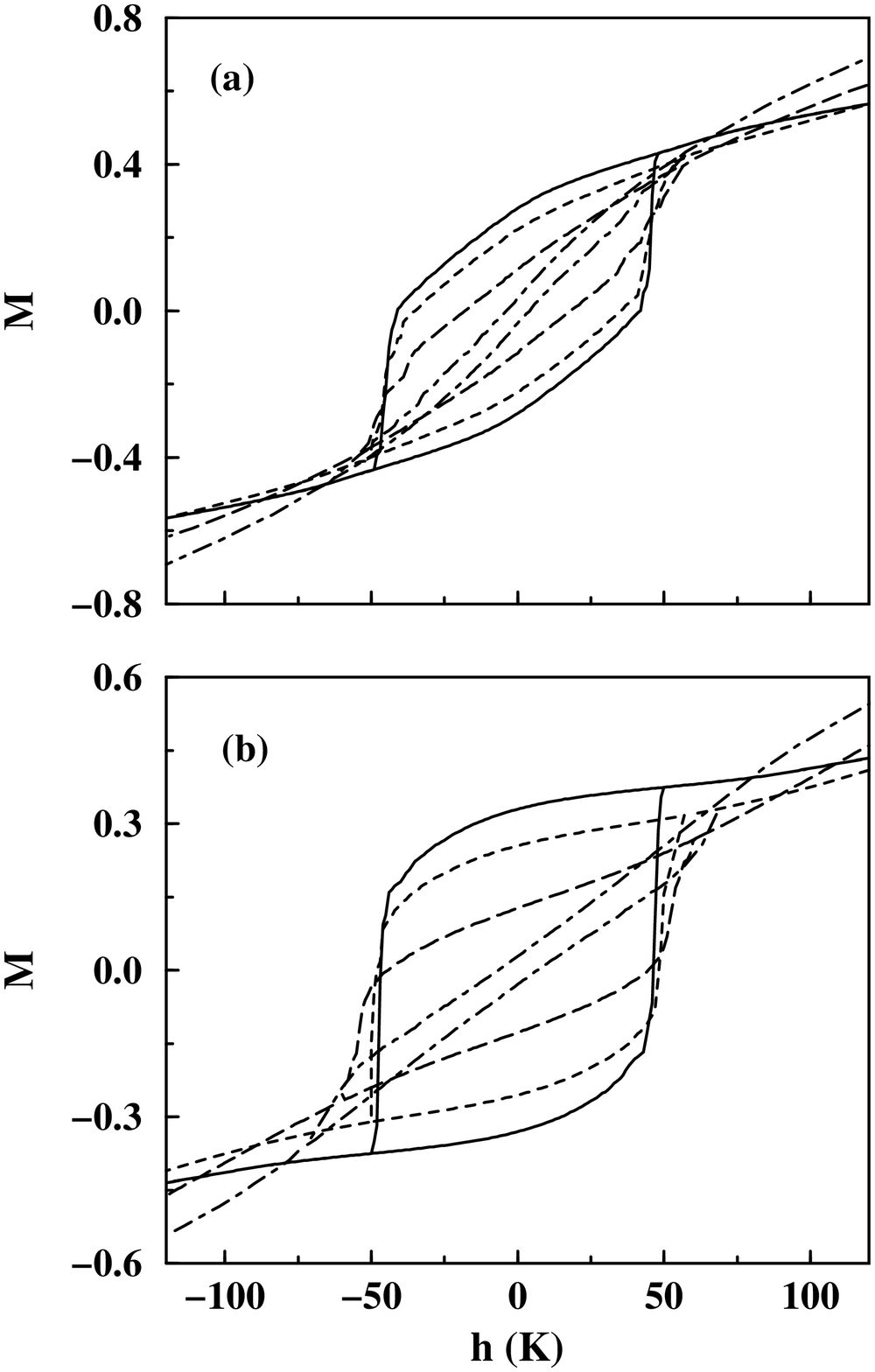}  
\caption{Hysteresis loops for systems with vacancy concentrations
$\rho_{v}$= 0.0, 0.166, 0.4, 0.6 (from outer to innermost) on the O sublattice at 
$T= 20$ K. Particle diameters $D=3$ (a), and $D=6$ (b). Results have been averaged 
over 10 disorder realizations.}  
\label{Fe2O3_HIDnvT20_fig} 
\efigu 
\bfigu[htbp] 
\centering 
\epsfxsize=13 cm  
\leavevmode  
\epsfbox{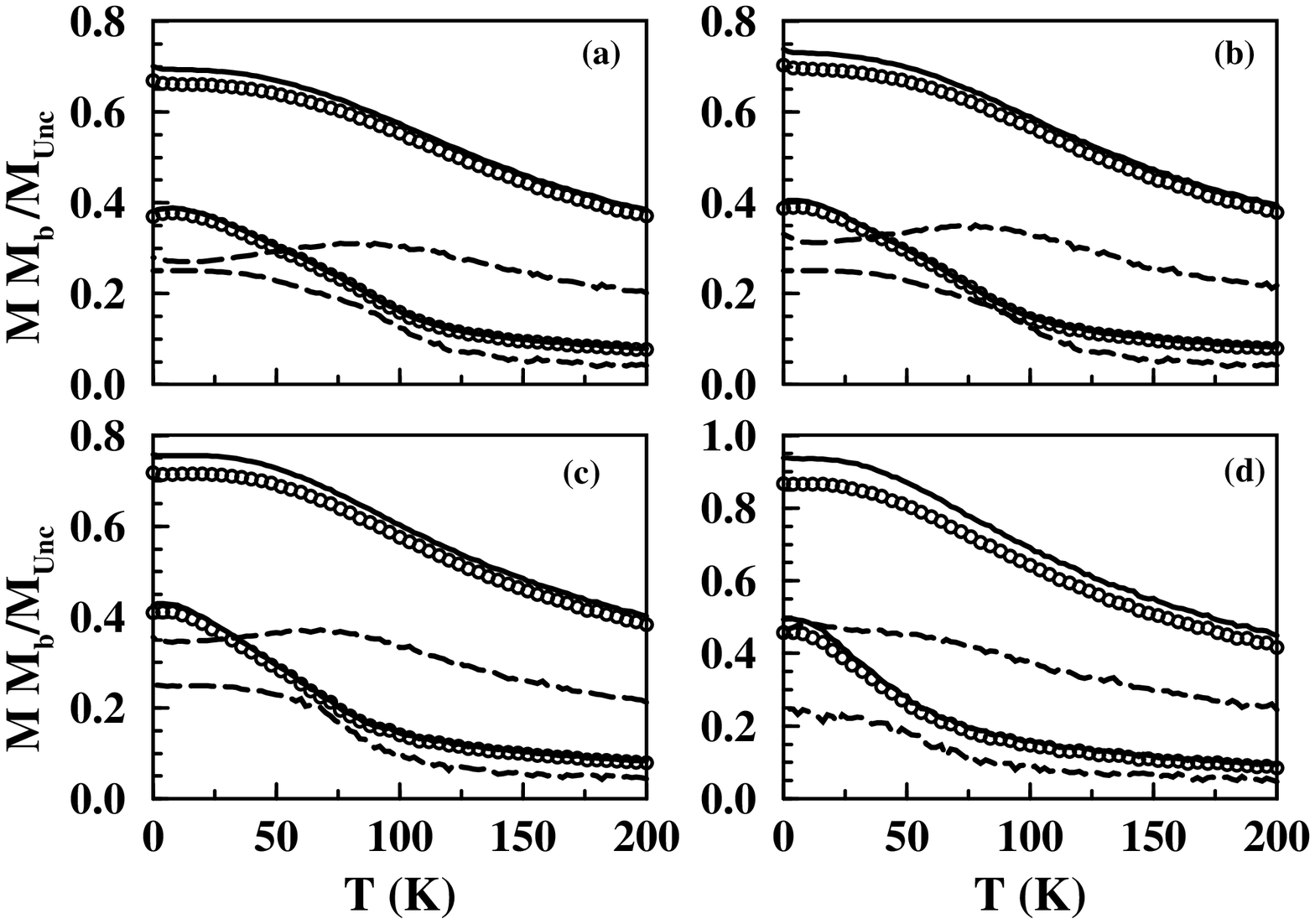}  
\caption{Thermal dependence of $M$ after cooling under a magnetic field 
for a spherical particle with $D= 3$, with vacancy densities 
on the surface of the O and T sublattices $\rho_{sv}=$ 0 (a), 0.1 (b), 0.2 (c), 
0.5 (d), and $\rho_{v}=$ 0.166 on the O sublattice.The results for 
two cooling fields $h_{FC}$= 20, 100 K (lower and upper curves respectively in 
each pannel) are shown. 
The contributions of the surface (thick lines) and the core (dashed lines) 
to the total magnetization (circles) have been plotted separatedly.  
The magnetization has been normalized to $M_{b}$, the magnetization of
a perfect ferrimagnetic configuration for a system of infinite size.}  
\label{Fe2O3_FCDsvD3_fig} 
\efigu 
\bfigu[htbp] 
\centering 
\epsfxsize=13 cm  
\leavevmode  
\epsfbox{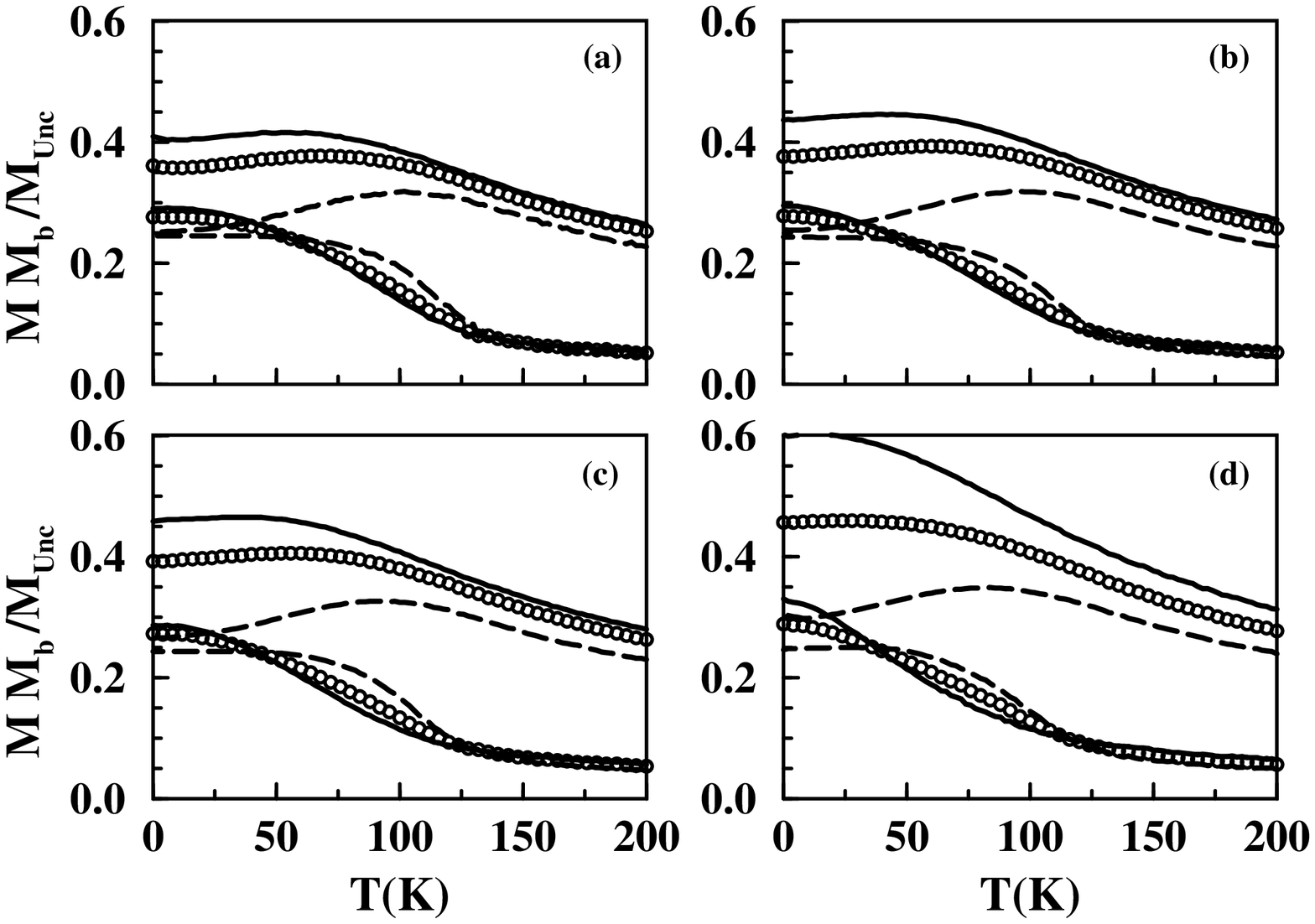}  
\caption{Same as Fig. 11 but for a spherical particle of diameter $D= 6$.}  
\label{Fe2O3_FCDsvD6_fig} 
\efigu
\bfigu[htbp] 
\centering 
\epsfxsize=13 cm  
\leavevmode  
\epsfbox{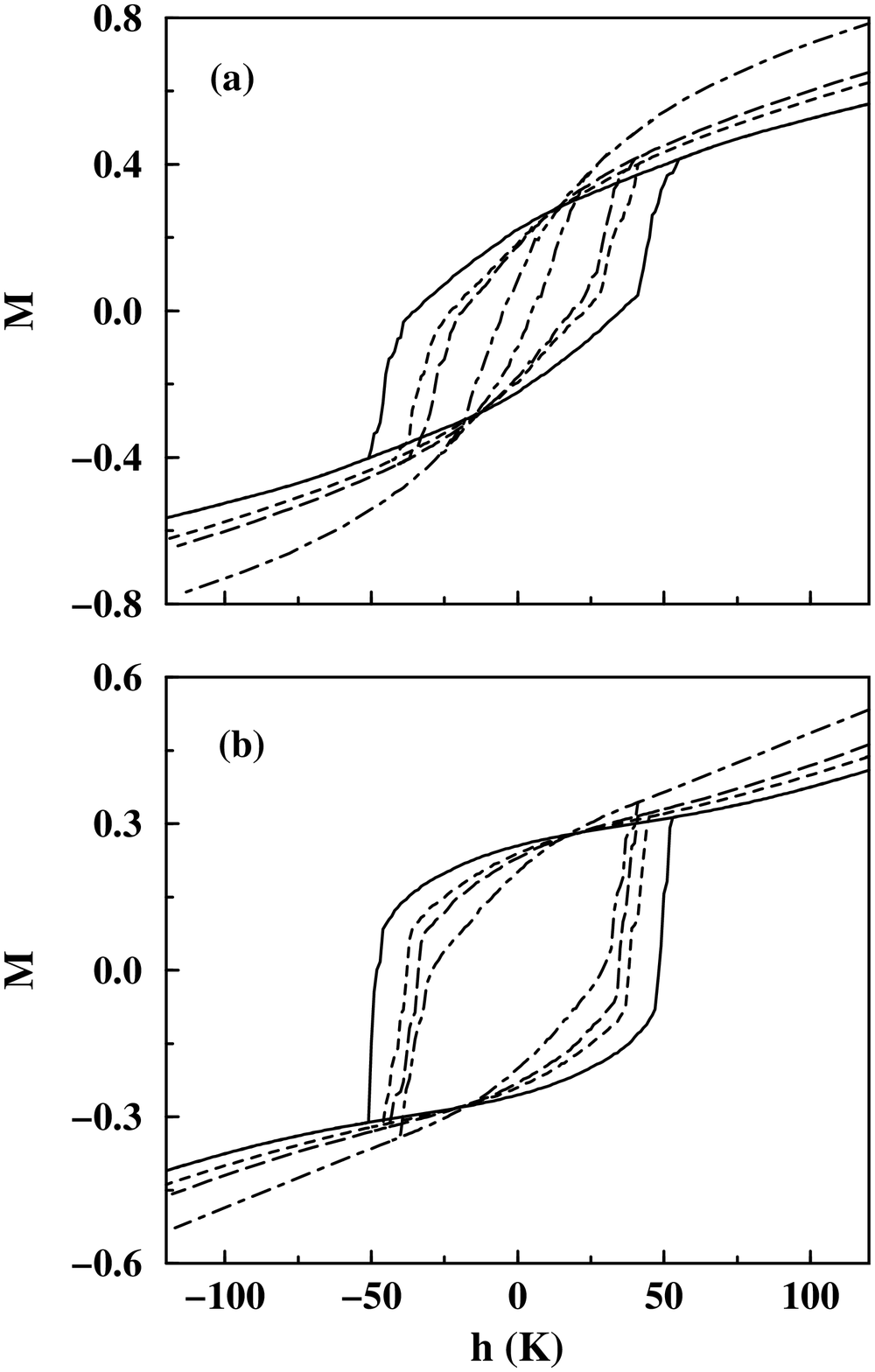}  
\caption{Hysteresis loops for systems with vacancy densities on the surface of the O 
and T sublattices $\rho_{sv}$= 0, 0.1, 0.2, 0.5, vacancy density 
$\rho_{sv}= 0.1666$ on the O sublattice, and $T= 20$ K.
Particle diameters $D=3$ (a), $D=6$ (b). 
Results have been averaged over 10 disorder realizations.}  
\label{Fe2O3_HIDsvT20_fig} 
\efigu 
\bfigu[htbp] 
\centering 
\epsfxsize=13 cm  
\leavevmode  
\epsfbox{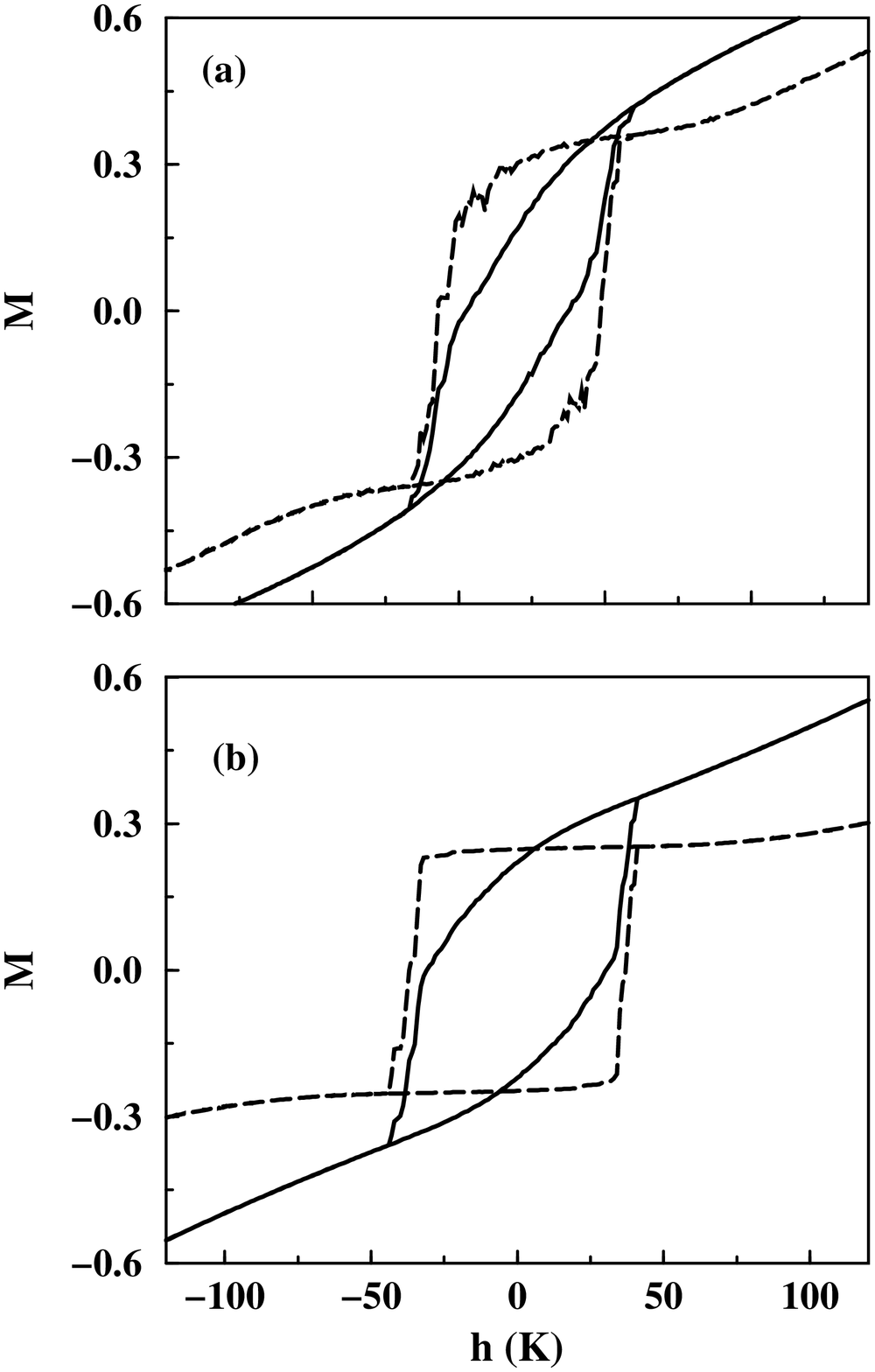}  
\caption{Core and Surface contributions for the case $\rho_{sv}= 0.2$ 
of Fig. 13.}  
\label{Fe2O3_HIDsvSC_fig} 
\efigu 
\bfigu[htbp] 
\centering 
\epsfxsize=13 cm  
\leavevmode  
\epsfbox{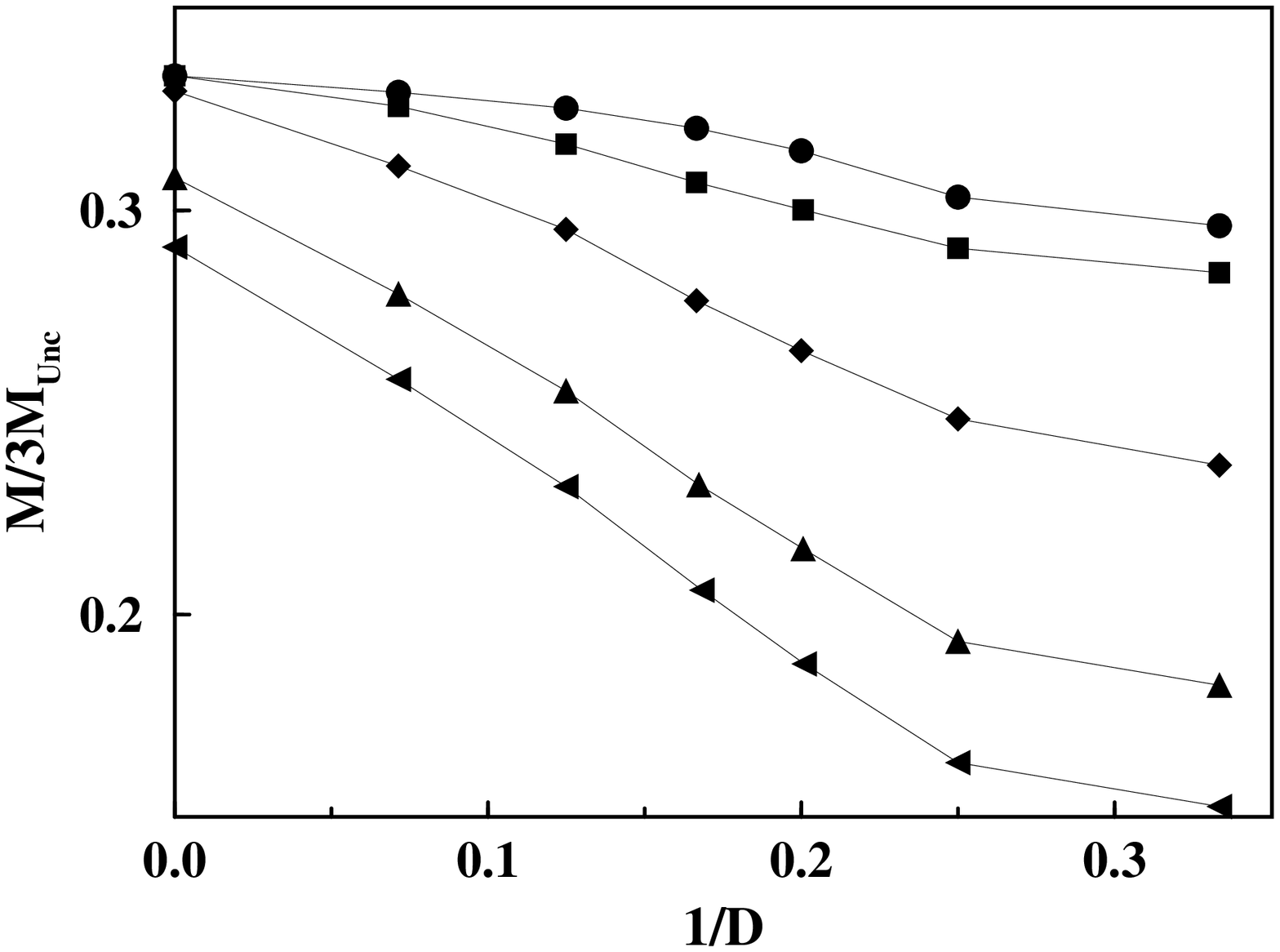}  
\caption{Size dependence of the magnetization of a spherical particle at 
different temperatures $T=$ 0, 20, 40, 60, 70 K (from upper to lowermost
curves).}  
\label{Fe2O3M(D)_fig} 
\efigu 
 
\end{document}